\input{aipcheck}

\documentclass[
    ,final            
  ]
  {aipproc}

\layoutstyle{8x11double}

\begin{document}

\title{The tip of the iceberg: the frequency content of the $\delta$ Sct star HD~50844
   from CoRoT space photometry}


\author{Poretti,~E.}{
  address={INAF-OA Brera, Via E. Bianchi, 46, 23807 Merate, Italy}
}

\author{Mantegazza,~L.}{
  address={INAF-OA Brera, Via E. Bianchi, 46, 23807 Merate, Italy}
}

\author{Rainer,~M.}{
  address={INAF-OA Brera, Via E. Bianchi, 46, 23807 Merate, Italy}
}
\author{Uytterhoeven,~K.}{
  address={INAF-OA Brera, Via E. Bianchi, 46, 23807 Merate, Italy}
  ,altaddress={Laboratoire AIM, CEA/DSM CNRS Universit\'e Paris Diderot, 91191, Gif-sur-Yvette, France} 
}

\author{Michel,~E.}{
  address={LESIA, Observatoire de Paris, CNRS (UMR 8109), 5 pl. Janssen, 92195 Meudon, France}
}

\author{Baglin,~A.}{
  address={LESIA, Observatoire de Paris, CNRS (UMR 8109), 5 pl. Janssen, 92195 Meudon, France}
}
\author{Auvergne,~M.}{
  address={LESIA, Observatoire de Paris, CNRS (UMR 8109), 5 pl. Janssen, 92195 Meudon, France}
}
\author{Catala,~C.}{
  address={LESIA, Observatoire de Paris, CNRS (UMR 8109), 5 pl. Janssen, 92195 Meudon, France}
}
\author{Samadi,~ R.}{
  address={LESIA, Observatoire de Paris, CNRS (UMR 8109), 5 pl. Janssen, 92195 Meudon, France}
}
\author{Rodr\'\i guez,~E.}{
  address={Instituto de Astrof\'\i sica de Andaluc\'\i a, Apartado 30040, 18080, Granada, Spain}
}
\author{Garrido,~R.}{
  address={Instituto de Astrof\'\i sica de Andaluc\'\i a, Apartado 30040, 18080, Granada, Spain}
}
\author{Amado,~P.}{
  address={Instituto de Astrof\'\i sica de Andaluc\'\i a, Apartado 30040, 18080, Granada, Spain}
}

\author{Mart\'\i n-Ruiz,~S.}{
  address={Instituto de Astrof\'\i sica de Andaluc\'\i a, Apartado 30040, 18080, Granada, Spain}
}
\author{Moya,~A.}{
  address={Instituto de Astrof\'\i sica de Andaluc\'\i a, Apartado 30040, 18080, Granada, Spain}
}

\author{Su\'arez,~J.C.}{
  address={Instituto de Astrof\'\i sica de Andaluc\'\i a, Apartado 30040, 18080, Granada, Spain}
}
\author{Baudin,~F.}{
  address={Institut d'Astrophysique Spatiale, CNRS, Universit\'e Paris XI UMR 8617, 91405, Orsay, France}
}
\author{Zima,~W.}{
  address={Instituut voor Sterrenkunde, K.U. Leuven, Celestijnenlaan 200 D, 3001 Leuven, Belgium}
}
\author{Alvarez,~M.}{
  address={Observatorio Astron\'omico Nacional,  UNAM, Apto Postal 877,
Ensenada, BC 22860, M\'exico}
}
\author{Mathias,~P.}{address={UMR 6525 H. Fizeau, UNS, CNRS, OCA, Campus Valrose, F-06108 Nice Cedex 2, France}
}

\author{Papar\'o,~M.}{
  address={Konkoly Observatory, PO Box 67, 1525 Budapest, Hungary}
}
\author{P\'apics,~P.}{
  address={Konkoly Observatory, PO Box 67, 1525 Budapest, Hungary}
}
\author{Plachy,~E.}{
  address={Konkoly Observatory, PO Box 67, 1525 Budapest, Hungary}
}
%

\begin{abstract}
It has been suggested that the detection of a wealth of very low
amplitude modes in $\delta$ Sct stars was only a matter of signal--to--noise
ratio. Access to this treasure, impossible from the ground, is one of the
scientific aims of the space mission CoRoT, developed and
operated by CNES. This work presents the results obtained on HD 50844: the
140,016 datapoints allowed us to reach the level 
of 10$^{-5}$ mag in the amplitude spectra.
The frequency analysis of the CoRoT timeseries
revealed hundreds of terms in the frequency range 0--30 d$^{-1}$. 
The initial guess that $\delta$ Sct stars
have a very rich frequency content is confirmed. The spectroscopic mode
identification gives theoretical support since very high--degree modes (up to
$\ell$=14) are identified. We also prove that cancellation effects are not
sufficient in removing the flux variations associated to these modes at the
noise level of the CoRoT measurements. The ground--based observations indicate
that HD 50844 is an evolved star that is slightly underabundant in heavy
elements, located on the Terminal Age Main Sequence. 
The predominant term ($f_1$=6.92 d$^{-1}$) has been identified
as the fundamental radial mode combining ground-based photometric and
spectroscopic data. 
\end{abstract}

\maketitle


\section{Introduction}
A huge effort has been made in the past decades to observe  $\delta$ Sct stars
from the ground. These opacity--driven pulsators show many excited
modes with amplitudes that can be reached by photometry from the ground. After several
results obtained by different teams performing single--site observations,
the most promising targets have been monitored by means of multisite campaigns.
A full description of the state--of--art in observing $\delta$ Sct stars
from ground can be found in Breger \& Montgomery \cite{vienna}.

The CoRoT\footnote{The CoRoT space mission was developed and  is operated by the French
space agency CNES, with participation of ESA's RSSD and Science Programmes,
Austria, Belgium, Brazil, Germany, and Spain.}
 (COnvection, ROtation and planetary Transits) space mission
gives us the possibility to investigate the pulsational content
of $\delta$ Sct stars from a totally new perspective, i.e., high--precision
measurements performed in a continuous way for tens of days (short runs)
or more than one hundred days (Long Runs).
One of the new $\delta$ Sct stars
discovered in the preparatory work, HD~50844, was among the 10 asteroseismologic targets of the 
Initial Run (IR01),
 started on February 2$^{\rm nd}$, 2007 and finished on March 31$^{\rm st}$, 2007
($\Delta$T=57.61~d).
It is a slightly evolved star, close
to the Terminal Age Main Sequence. The physical parameters derived from Str\"omgren
photometry are $M_V$=1.31, T$_{\rm eff}$=7500~K, $\log g$=3.6, and [Fe/H]=--0.4~dex
\citep{anti}. In this contribution we summarize the results obtained from the analysis
of the CoRoT data and the ground--based spectroscopy and photometry; Poretti et al. \cite{main} give a more detailed
description of all the procedures and discussions sketched here.  

\begin{figure*}
  \includegraphics[width=1.0\textwidth]{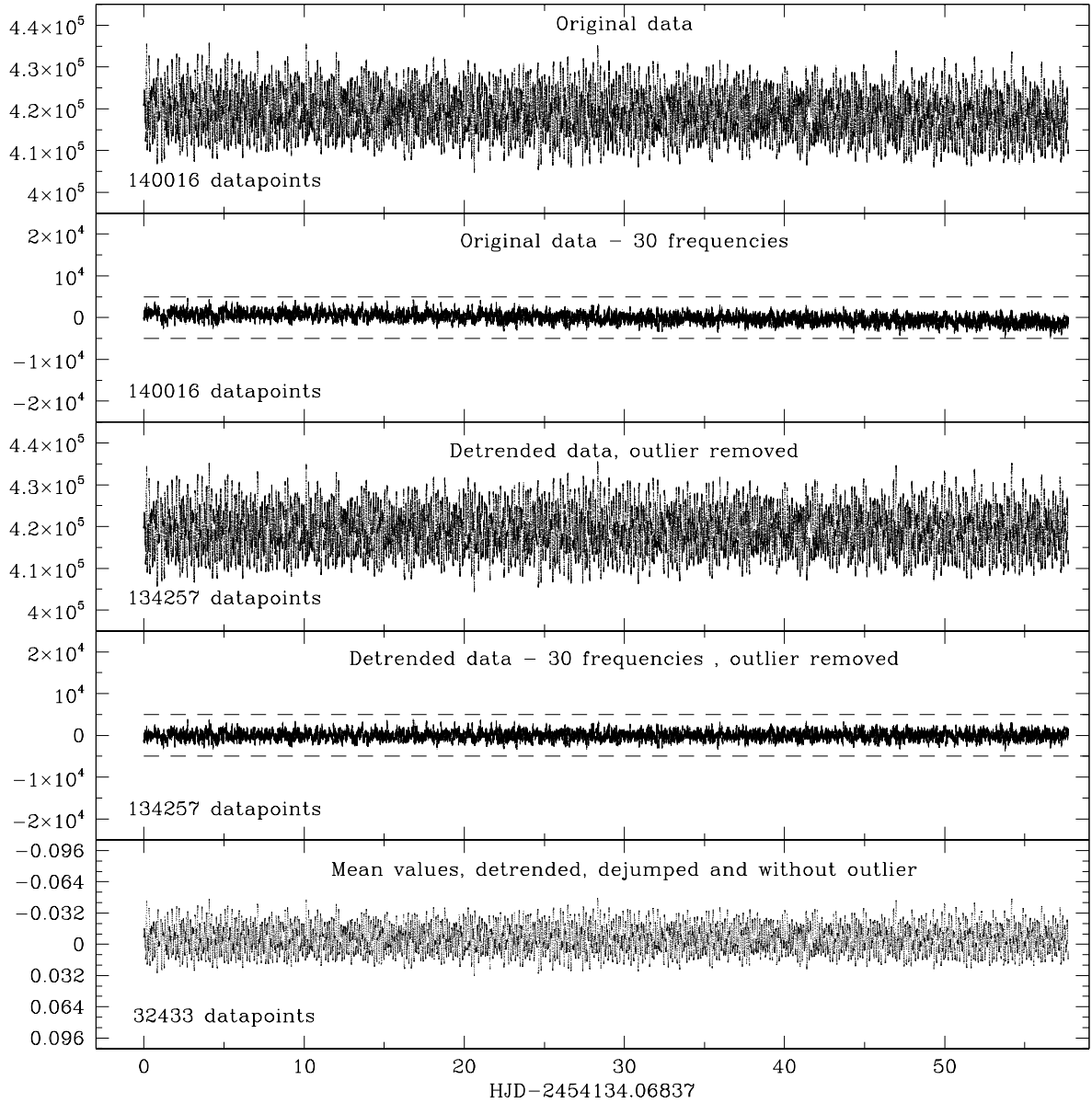}
  \caption{The different steps in preparing the final dataset of the CoRoT measurements
of HD~50844. The measure units of the y--axes are observed flux for the first and third
panels, residual flux for the second and fourth panels, magnitudes for the fifth panel. }
\label{curve}
\end{figure*}

\section{The reduction of the CoRoT data}

For the analysis, we used the reduced N2 data rebinned at 32~sec
and  we only
considered the 140,016 datapoints for which no problem (i.e., flag=0) were reported
(Fig.~\ref{curve}, top panel).
The rejection of the flagged points resulted in a slight
enhancement of the amplitude  of the orbital frequency, since the bad measurements
occurred mostly when the satellite  crossed the South Atlantic Anomaly (SAA).
The light curve  was detrended with a linear fit to remove the effect of ageing
\citep{flight}. Firstly, the thirty frequencies with the highest amplitudes were
identified and the data were prewhitened to clearly show outliers, jumps and long--term
trend (Fig.~\ref{curve}, second panel). The final dataset is composed of
134,257 datapoints (Fig.~\ref{curve}, third panel) and the subsequent prewhitening leaves a residual light curve
without appreciable trend (Fig.~\ref{curve}, fourth panel).
To gain in CPU time and to reduce the noise level, in the last step we grouped the original data
into new bins of four  consecutive measurements, thus
obtaining 32,433 datapoints (Fig.~\ref{curve}, bottom panel).

\begin{figure}
  \includegraphics[width=0.48\textwidth]{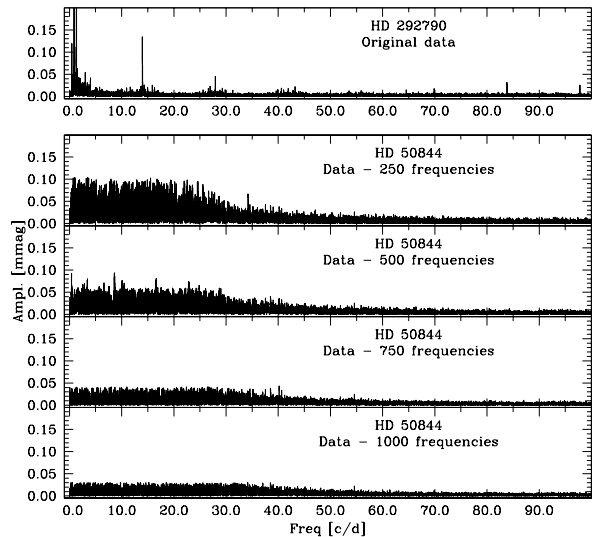}
  \caption{For comparison purposes, the top panel  shows the amplitude spectrum of the rotational variable
HD~292790. This star has been observed by CoRoT simultaneously
to HD~50844. The amplitude spectra of HD~50844 shown
in the other panels indicate the extreme richness of the pulsational spectrum
of this $\delta$ Sct star: up to 1000  peaks  are needed to make the  residual
spectrum of HD~50844 comparable to the original one of HD~292790.
}
\label{spectrum}
\end{figure}

To have a reference frame for interpreting the results for HD~50844 (A2, $V$=9.09),
we considered the data of HD~292790 (F8, $V$=9.48), which  was observed by CoRoT  in the same IR01.
Our frequency analysis shows that the luminosity of HD~292790 is modulated by the rotation in
a  simple way.
The amplitude spectrum shows the rotational frequency,
its harmonics and the satellite frequencies (Fig.~\ref{spectrum}, top panel).
For most part of the  spectrum, i.e., from 10 to 100~d$^{-1}$, the
noise level is distributed in an uniform way and is very  low, namely below 0.01~mmag.
For $f<5~$d$^{-1}$, where the modulation terms and the long term
drift are concentrated, the noise level slightly increases.
On the other hand, the amplitude spectrum of HD~50844 appears to be very dense 
(Fig.~\ref{spectrum}, second panel).
At first glance, it is clear that the amplitude spectrum of
HD~50844 is only for $f>50~$d$^{-1}$ as flat as that of HD~292790.
We can calculate a noise  amplitude of $7.5\times10^{-6}$ mag.
It is straightforward to deduce that the signal is concentrated in the
$f<30~$d$^{-1}$ domain, though there is no predominant peak
region after identification of  250 frequencies (residual rms 1.2 mmag).
At that point, the  average peak height (0.10~mmag) for $f<30~$d$^{-1}$ is still 10
times higher than for $f>50~$d$^{-1}$.
A huge number of terms is
necessary to reduce the amplitude spectrum of HD~50844 at the expected noise level.
The peak height gets  progressively lower after the identification of
500, 750, and 1000 terms (third, fourth,  
and fifth panels in Fig.~\ref{spectrum}, respectively).
It is possible that
some of the lower peaks originate from changes in amplitude or phase, or  from other
effects intrinsic to the star. In any case,
the main conclusion of our analysis and related checks is that hundreds of terms
--~at least up to 1000~--
are needed  to explain the light variability of HD~50844.
The detection of so many independent terms in the light variability of a
 $\delta$ Sct star is a totally new result.

\begin{figure}
  \includegraphics[width=0.48\textwidth]{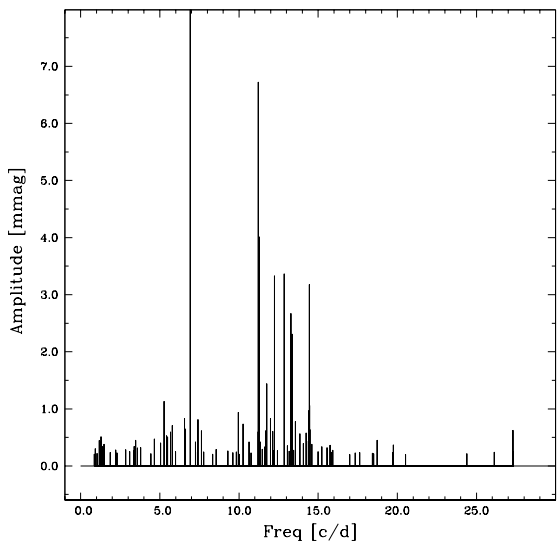}
  \caption{The amplitude spectrum of HD 50844, limited to the first 100 frequencies.
The amplitude of the predominant term at 6.92~d$^{-1}$ is out of scale (15.4 mmag).}
\label{distri}
\end{figure}

Figure~\ref{distri} shows the distribution of the first 100 frequencies identified
in the CoRoT timeseries.




\section{Mode indentification from ground--based high-resolution spectroscopy and  multicolour
photometry}

The spectroscopic observations were completed from  January 2 to
28, 2007
with the FEROS instrument mounted at the {2.2-m} ESO/MPI telescope,
La Silla, Chile.
The spectral resolution of FEROS is $R\sim 48,000$. We obtained 232 spectra
in 14  nights. The exposure time was set to 900~sec and
the signal-to-noise ratios (SNRs) ranged from 150 to 210. The spectra were reduced
using an improved version of the standard FEROS pipeline, written in MIDAS and
developed by Rainer \cite{monica}.
The mean barycentric radial velocity of the star is $-10.8\pm$0.2~km~s$^{-1}$.
We also derived  a $v\sin~i$ value of 58$\pm$2~km~s$^{-1}$.

The spectroscopic frequencies were searched for by analysing the LPVs
by means of the pixel--by--pixel technique \citep{manvie,zima}.
We detected 27 terms with an SNR higher than 4.0 and 3 terms with 3.1$<$SNR$<$3.5.
The mode identification
was performed by fitting the amplitude and phase variations of each mode across the line profile
\citep{manvie} by using the software FAMIAS \citep{famias}.
Uncertainties
are estimated to be  $\pm$1 for the degree  $\ell$ and $\pm$2 for the order  $m$.

The main goal of the spectroscopic observations
is to answer the crucial question about the huge number of
 frequencies detected in the CoRoT photometric data:
is the large number of photometric terms related to the visibility not only of low--degree modes but
also of high--degree modes?
To investigate this we focus on the mode identification results, as derived from the high-resolution data.
The highest detected degree is  $\ell=14$. This observational fact tells us that
we have at least 235 possible modes of pulsation for a given radial order $n$.
 A few  different radial orders are
sufficient for explaining the 1000 frequencies detected in the CoRoT photometric
timeseries of HD~50844.
On the other hand,
we know from CoRoT photometry that the frequency values higher than 30~d$^{-1}$
are observed after the 300$^{\rm th}$
rank of detection, i.e.,  with very low amplitudes. The observed spectroscopic
frequencies have $f<$20~d$^{-1}$.  We still miss the spectroscopic
counterpart of the {$20<f<50$~d$^{-1}$} region. Such a region could be
filled by  modes with high radial orders $n$, which are shifted
toward higher frequencies than those of the low orders.

In addition to the LPV analysis, the
FEROS spectra  were used to determine the  abundances of the
elements. HD~50844 is a slightly metal-deficient star. However,
the abundances of elements as C, N, and S
in the HD~50844 atmosphere are very similar to those of the Sun, while
other elements are underabundant.  Such an
abundance pattern is  typical of $\lambda$ Boo stars \citep{paunzen}.

The importance of colour information to estimate the degree $\ell$
of the modes is   well-known  \citep{garvie}; therefore,
we observed HD~50844 in Stromgren $uvby$ photometry
to accompany the CoRoT white--light photometry.
Several campaigns were carried out
at S. Pedro M\'artir and Sierra Nevada observatories, using  twin
Danish photometers.
We calculated the amplitude ratios and the phase shifts  by using
the $y$ colour as the reference system.
The immediate result is that the predominant term $f_1$=6.92~d$^{-1}$
is characterized by positive phase shifts (i.e., $\phi_{u,v,b}-\phi_y>0.0$).
The mode identification of the $f_1$ term from the spectroscopic data was not trivial since we
had to take the harmonic 2$f_1$ and the equivalent width variations into account.
The FAMIAS method returns two possibile $(\ell,m)$ couples, i.e.,
(0,0) and (2,0). When considering both the spectroscopic and the photometric identifications,
we can consider the $f_1$ term as the fundamental radial mode.

\section{Conclusions}
The exploitation of the CoRoT photometric timeseries of HD~50844 resulted in
a very complex and intriguing task, also giving a completely new picture of the
pulsational content of $\delta$ Sct stars. We demonstrated that
the light curve of HD~50844 can be explained by the presence of
hundreds of excited terms. Classical checks such as using  different
software packages, subdividing the timeseries into different subsets,
carefully inspecting the residuals, and  comparing with other similar targets
observed by CoRoT, together confirm this issue. High--resolution spectra support
the excitation of modes having a very high degree $\ell$ (up to $\ell$=14),
thus providing an observational explanation  for the richness of the frequency spectrum.

Such a large number of modes of different $\ell$ simultaneously excited is among plausible
features suggested by theoretical works.
An immediate conclusion is that the cancellation effects are not sufficient
in removing the variations of the flux integrated over the whole stellar disk;
Daszy\'nska-Daszkiewicz et al. \cite{romapol} correctly predicted this extension
to high degrees of the  modes
extracted from  high--precision photometric timeseries.
We should  keep in mind that the CoRoT performances and  the continuous
monitoring allow  detection of amplitudes of about 10$^{-5}$~mag without significant
aliasing effects, which is  a totally new perspective.

The case of HD~50844 seems to match the theoretical picture
featuring a very large number of excited modes with amplitudes limited by a saturation
mechanism of the $\kappa$ mechanism.
This confirms that the observation of $\delta$ Sct stars with CoRoT will allow us to
address the longstanding problem of amplitude distribution and the potential
existence of mode selection processes in the $\delta$ Sct instability strip.
We note that HD~50844  probably does not constitute a ``standard" scientific case
of a $\delta$ Sct star, because it is
located on the TAMS,  probably belongs to the class of $\lambda$ Boo stars (i.e.,
showing atmospheric particularities), and it is  seen almost equator-on.
Nevertheless, seismic models
that were able to reproduce the position of HD~50844 in the HR
diagram and its measured rotational velocity were calculated.

HD~50844 has for the first time disclosed the intrinsic complexity
of  frequency spectra of intermediate--mass stars in the lower part of the
instability strip.
It is quite evident that, after these first CoRoT observations, we must look at $\delta$ Sct
stars in a completely new way. We have to realize that ground--based observations were
only able to observe the tip of the iceberg, while the larger part of the modes have
remained undetected.


\begin{theacknowledgments}
The FEROS data are being obtained as part of the ESO Large Programme
LP178.D-0361 (PI.: E.~Poretti). 
Mode identification results were
obtained with the software package FAMIAS developed in the framework of the
FP6 Coordination Action Helio- and
Asteroseismology (HELAS; http://www.helas-eu.org/). WZ was
supported by the FP6 European Coordination Action HELAS and by the Research
Council of the University of Leuven under grant GOA/2008/04.
This work was supported by the Italian
ESS project, contract ASI/INAF I/015/07/0, WP 03170, and by the Hungarian
ESA PECS project No. 98022. KU acknowledges financial support from a
{\it European Community Marie Curie Intra-European Fellowship},
contract number MEIF-CT-2006-024476. 
JCS acknowledges support from the {\it Consejo Superior de Investigaciones
Cient\'{\i}ficas} by an I3P contract financed by the European Social Fund
and from the Spanish {\it Plan Nacional del Espacio} under project
ESP2007-65480-C02-01.  PJA acknowledges financial support
from a {\it Ramon y Cajal} contract of the Spanish Ministry of Education
and Science. AM ackowledges financial support from a {\it Juan de la Cierva}
contract of the Spanish Ministry of Science and Technology.
SMR acknowledges a {\it Retorno de Doctores} contract financed by the Junta de Andaluc\'{\i}a and
Instituto de Astrof\'{\i}sica de Andaluc\'{\i}a for carrying out observing campaigns for CoRoT targets
at Sierra Nevada Observatory.
EN acknowledges financial support of the N~N203~302635
grant from the MNiSW.

\end{theacknowledgments}



\bibliographystyle{aipprocl} 

\begin{thebibliography}{9}
\bibitem{flight} Auvergne, M., Bodin, P., Boisnard, L., et al.,
2009, A\&A, in press (arXiv:0901.2206)
\bibitem{vienna}
Breger, M., Montgomery, M.H.., Editors, \emph{Delta Scuti and Related Stars},
ASP Conf. Series, vol.~210
\bibitem{romapol} Daszy\'nska-Daszkiewicz, J., Dziembowski, W.A.,
\& Pamyatnykh, A.A., 2006, Mem. SAIt, vol.~77, 113
\bibitem{garvie} Garrido, R., 2000,
in ``Delta Scuti and Related Stars", M.~Breger \& M.H.~Montgomery Eds., ASP Conf. Series, 210, 67
\bibitem{manvie} Mantegazza, L., 2000,
in ``Delta Scuti and Related Stars", M.~Breger \& M.H.~Montgomery Eds., ASP Conf. Series, 210, 138
\bibitem{paunzen} Paunzen, E., 2004, in ``The A-Star Puzzle",
J.~Zverko, J.~Ziznovsky, S.J.~Adelman, \& W.W.~Weiss Eds., Proc. IAU Symp. 224
(Cambridge University Press), p.~443
\bibitem{anti} Poretti, E., Alonso, R., Amado, P.J., et al., 2005, AJ, 129, 2461
\bibitem{main} Poretti, E., Michel, E., Garrido, R., et al., 2009, A\&A in press
(arXiv:0906.2628)
\bibitem{monica} Rainer, M., 2003, Laurea Thesis (in Italian), Universit\`a degli
Studi di Milano
\bibitem[Zima(2006)]{zima} Zima, W., 2006, A\&A, 455, 227
\bibitem[Zima(2008)]{famias} Zima, W., 2008, CoAst, 155, 17
\end{thebibliography}

\end{document}